\documentclass[mathleft]{an}
\usepackage{graphicx}
\usepackage{subfigure}
\usepackage{times}
\begin{document}

\Pagespan{789}{}
\Yearpublication{2017}%
\Yearsubmission{2017}%
\Month{11}%
\Volume{999}%
\Issue{88}%

\title{Could the low braking index pulsar \\ PSR J1734$-$3333 evolve into a magnetar?}

\author{Z.-F. Gao\inst{1,2}\and D.-L. Song\inst{3}\and Y.-L. Liu\inst{4,1}\and X.-D. Li\inst{5}\and N. Wang\inst{1}\fnmsep\thanks{Corresponding author:
  \email{na.wang@xao.ac.cn}\newline}\and H. Shan\inst{1}}

\titlerunning{The magnetic field evolution of PSR J1734$-$3333}
\authorrunning{Gao et al.}
\institute{Xinjiang Astronomical Observatory, CAS,150, Science 1-Street, Urumqi, Xinjiang, 830011, China
\and Key Laboratory of Radio Astronomy, Chinese Academy of Sciences, West Beijing Road, Nanjing, 210008, China
\and The Information Engineering University, 62 Science Road,
ZhengZhou, Henan, 450001, China
\and School of Physics, China West Normal University, Nanchong, Sichuan, 637002, China
\and Department of Astronomy and Key Laboratory of Modern Astronomy and Astrophysics, Nanjing University, Jiangsu 210046, China}

\received{12 August 2017}
\accepted{22 August 2017}
\publonline{later}
\abstract{ The low braking-index pulsar PSR J1734$-$3333 could be born with superhigh internal magnetic fields $B_{\rm in}\sim10^{15}-10^{16}$ G, and undergo a supercritical accretion soon after its formation in a supernova explosion. The buried multipole magnetic fields will merger into a dipole magnetic field. Since the magnetic flow transfers from the core to the crust of the pulsar, its surface dipole field grows quickly at a power-law form assumed until it saturates at the level of internal dipole field. The increase in surface dipole magnetic field results in the observed low braking index of $n=0.9(2)$. Keeping an average field growth index $\varepsilon=1.34(6)$, this pulsar will become a magnetar with surface dipole magnetic field at the equator $B_{\rm d}\sim 2.6(1)\times 10^{14}$\,G and $\sim 5.3(2)\times 10^{14}$\,G after next 50\,kyrs and 100\,kys, respectively.}
\maketitle

\section{Introduction}
Pulsars are among the most mysterious objects in the universe
that provide natural laboratory for investigating the nature of matter
under extreme conditions\,(e.g., Graber et al. 2015, 2017; Lai \& Xu 2016; Dong et al. 2013, 2016; Liu 2016; Liu et al. 2016; Mu et al. 2017; Xia \& Zhou 2017; Zhao 2015, 2016; Zhou et al. 2017). There are several methods for roughly measuring the magnetic field strength of a pulsar, such
as magneto-hydrodynamic pumping, Zeeman splitting,
cyclotron lines, magnetar bursts and etc\,(e.g., Cheng et al. 2013, 2015; Weng \& Zhang 2015; Weng et al. 2017; Liu et al. 2017; Liu \& Liu 2017).
 The surface diploe magnetic field of a pulsar can be estimated by measuring its spin period $P$ and period derivative $\dot{P}$. If the magnetic dipole radiation\,(MDR) solely causes the pulsar to spin down, the diploe magnetic field at the magnetic equator $B_{\rm d}$ is inferred as
\begin{equation}
\label{1}
P\dot{P} = bB_{\rm d}^2
\quad \Leftrightarrow \quad
B_{\rm d}= 3.2\times 10^{19}\mbox{ G }(P\dot{P})^{1/2},
\end{equation}
where $b=8\pi^{2}R^{6}\sin^{2}\alpha/3c^{3}I$,
$I$ and $R$ are the moment of inertia and radius of a NS, respectively,
$\alpha$ is the angle between the stellar rotation and magnetic axes.

Magnetars are a kind of special
pulsars powered by the decay of their strong magnetic fields (e.g., Gomes et al. 2014, 2017; Tauris \& Konar 2001) and identified as anomalous X-ray pulsars (AXPs) or soft gamma repeaters (SGR). The most remarkable feature of magnetars is their violent outbursts, during which the X-ray
luminosity can increase by a few orders of magnitude.

PSR J1734$-$3333 is a high magnetic field pulsar with an inferred dipole magnetic field $B_{\rm d}=5.2\times 10^{13}\mbox{ G }$.
PSR J1734$-$3333 is a high magnetic field pulsar with $B_{s}=5.2\times 10^{13}\mbox{ G }$, the period $P =1.17$ s, the period derivative $\dot{P}= 2.28\times10^{12}$\,s\,s$^{-1}$, and the second period derivative $\ddot{P}=(5.0\pm0.8)\times10^{-24}
$\,s~s$^{-2}$, respectively. It is located between the normal radio pulsars and magnetars in the $P-\dot{P}$ diagram. PSR J1734$-$3333 has been observed regularly since 1997 by using the 64\,m telescope at Parkes and the 76m telescope at Jodrell Bank. It has not glitched during these years, insuring the accuracy of measuring $\nu$, $\dot{\nu}$ and $\ddot{\nu}$, where $\nu=1/P$ is the pulsar rotation frequency,\,$\dot{\nu}$ and $\ddot{\nu}$, are its first and second derivatives, respectively.

The braking index $n$ of a pulsar describes the
dependence of the braking torque on rotation frequency\,(e.g., Lyne et al. 1993). The standard way to define the braking index is
\begin{equation}
\label{2}
  n=\frac{\Omega\ddot{\Omega}}{\dot{\Omega}^{2}}=\frac{\nu\ddot{\nu}}{\dot{\nu}^{2}} =2-\frac{P\ddot{P}}{\dot{P}^{2}},
\end{equation}
where $\Omega$, $\dot{\Omega}$ and $\ddot{\Omega}$ are the angular velocity, the derivative and the second derivative of $\Omega$, respectively, $\nu$ is the rotation frequency, $\dot{\nu}$ and $\ddot{\nu}$ are its first and second derivatives,
respectively.

As we know, the braking index of a pulsar is determined by its slow down
torque. If the magnetic fields of pulsars are constant, the MDR model
predicts the braking index $n=3$. Recently, Magalhaes et al.\,(2012, 2016) have modified canonical model to explain the observed braking indices ranges. Other models were also proposed to explain the low braking indices of pulsars\,(e.g., Menou et al. 2001; Chen \& Li 2016; Dupays et al. 2008, 2012; Coelho et al. 2016). The gravitational wave\,(GW) radiation\,(Cheng et al. 2017a, 2017b; de Araujo et al. 2016), the dipole magnetic field decay and/or the magnetic inclination angle decrease can interpret the higher braking index of PSR J1640$-$4631\,(e.g., Ek\d{s}i et al. 2016; Gao et al. 2017).

The present braking theories of pulsars are challenged by the relatively small braking index of PSR J1734$-$3333. Assuming
that the magnetic field and the particle luminosity are
both constants, this source may be braking by a rotation powered
particle wind from magnetosphere\,(e.g., Kou \& Tong 2015; Tong \& Kou 2017; Kou et al. 2016; Yuen \& Melrose 2014, 2017). Adopting a modified formula for the propeller torque, a self-similar fall-back disk can account for the small braking index, $P$, $\dot{P}$ of PSR J1734$-$3333 (Liu et al. 2014). Other models\,(e.g., Chen \& Li 2016; Ertan et al. 2008) were  proposed, but these models cannot explain magnetar-like bursts from this pulsar.

Here we consider multipole magnetic fields buried soon after birth and diffuse to the surface. By combining the dipole magnetic field growth with the MDR model, we will investigate the evolution of PSR J1734$-$3333.  Recently, magnetar-like outbursts from PSR J1734$-$3333 were reported\,(e.g., G\"{o}\v{g}\"{u}\c{s} et al. 2016). These outbursts could be caused by the decay of initial multipole magnetic fields. The multipole magnetic fields are merging through crustal tectonics to form dipole magnetic field, which causes a growth in the surface dipole magnetic field.

We assume that the internal multipole magnetic fields $B_{\rm in}\sim 10^{15-16}$\,G anchored in the inner crust and extend to the core and the surface dipole field $B_{\rm d}$ after the NS formation, but prior to mass accretion.  Accretion then buries and compresses these born multipole and dipole magnetic fields, and the surface dipole field at birth is far less than the quantum critical magnetic field.
The buried multipole magnetic fields will merger and the buried dipole magnetic field will diffuse via the Hall drift and Ohmic decay on a much longer timescale, $\tau >10^{6}$ yr, especially if the core is super conducting. Since the magnetic flow transfers from the core to the crust of the pulsar, its surface dipole field $B_{\rm d}(t)$ grows quickly. The increase in $B_{\rm d}(t)$ may result in a small braking index of about $n\sim 0.9$\,(Espinoza et al. 2011). By matching the age of PSR~J1734$-$3333, we should constrain the pulsar's dipole magnetic field and spin period at birth and investigate whether it can obtain magnetar-like magnetic field strength in the future field evolution.

\section{The standard scenario for magnetic field growth in PSR~J1734$-$3333  }

The standard scenario for the rotational evolution of a pulsar is that it is born rapidly spinning and rapidly slowing or spinning down, i.e., large $\dot{P}$. A newborn pulsar would be placed in the top-left region of $P-\dot{P}$ diagram. As it spins down, the pulsar moves for $\sim 10^5-10^6\mbox{yr}$ toward the bottom-right of $P-\dot{P}$ diagram\,(Ho 2015). \begin{equation}
\label{3-equation}
\frac{\partial \vec{B}}{\partial t}=-\nabla\times\left[\frac{c^{2}\nabla}{4\pi\sigma}\times e^{\nu}\vec{B}+\frac{c\nabla}{4\pi en_{e}}\times e^{\nu}\vec{B}\times\vec{B}\right],
\end{equation}
where $\sigma$ is the electric conductivity parallel
to the magnetic field, $e^{\nu}$ is the relativistic
red-shift correction, $n_{e}$ is the electron number density,
and $e$ the electron charge.
This equation contains two different effects that act on two distinct timescales, which can be estimated as
\begin{equation}
\label{4-equation}
t_{\rm Hall}=\frac{4\pi n_{e}e L^2}{cB},~~~ t_{\rm Ohm}=\frac{4\pi \sigma L^{2}}{c^{2}},
\end{equation}
where $t_{\rm Ohm}$ is the Ohmic dissipation timescale with a typical value of $\sim 10^{6}$\,yrs or more\,(e.g., Muslimov \& Page 1996; Vigan\`{o} et al. 2013),
$t_{\rm Hall}$ is the Hall drift timescale with a typical value of several $\times(10^{4}-10^{5})$\,yrs for high magnetic field pulsars and magnetars, and $L$ is a characteristic length scale of variation, which can be taken to be the thickness of the neutron star crust\,(Ho 2011).

The buried magnetic field of PSR J1734$-$3333 were assumed to evolve as
\begin{equation}
\label{5}
B(t) = \frac{B_0}{1+\,t/\tau_{\rm D}},
\end{equation}
where $\tau_{\rm D}$ is the effective field decay timescale which is
approximately equal to $\tau_{\mathrm{Ohm}}$ for radio pulsars or $\tau_{\mathrm{Hall}}$ for magnetars. The observed low braking indices can be attributed to an increase in dipole magnetic field: allowing $B$ in Equation (2) evolve, one obtains
\begin{equation}
\label{6}
n = 3 - 2\frac{\dot{B_{\rm d}}}{B_{\rm d}}\frac{P}{\dot{P}}
 = 3 - 4\tau_{\rm c}\frac{\dot{B_{\rm d}}}{B_{\rm d}}
\end{equation}
where $\tau_{\rm c}=2P/\dot{P}$ is the characteristic age for the star. In order to bury the magnetic field at a particular density, the field must be buried at a greater depth for a lower $M$. Using an age estimate of $t>1.3$\,kyrs\,(Ho \& Andersson 2012), Ho (2011, 2015) constrained the initial magnetic field strength to $B_0\sim (1-3)\times10^{14}\mbox{ G}$, the initial period $P_0$ to 100 ms for the pulsar. According to their numerical results, the NS crust thickness decreases with increasing mass for a given nuclear equation of state\,(e.g., Potekhin et al. 2013).

\section{Alternative scenario for magnetic field growth in PSR~J1734$-$3333 }
\subsection{Surface dipole magnetic field growth}
In previous studies on the magnetic field evolution of NSs, the magnetic fields constrained in the crust of the star are interested\,(e.g., Zhang \& Xie 2012; Potekhin et al. 2013; Gomes et al. 2017; Liu et al. 2016a, 2016b, 2017). Here we assume that PSR~J1734$-$3333 has magnetar-like internal magnetic fields\,(Gao et al. 2014, 2016)\,$B_{\rm in}\sim 10^{15}-10^{16}$\,G at the birth. These fields are constrained in the inner crust and the core after an early episode of accretion and are slowly diffusing to the surface.  We consider the evolution of the dipole magnetic field in the NS crust-core. Neglecting the hydrodynamic motions, thermomagnetic effects, and anisotropy of the electrical conductivity, the magnetic stream function $S=S(r, t)$ follows
\begin{equation}
\label{7}
 \frac{\partial S}{\partial t}= \frac{c^2}{4\pi\sigma}\left(\frac{\partial^{2}S}{\partial r^{2}}-\frac{\partial S}{r^{2}}\right).%
\end{equation}
At the surface $(r=\,R)$, the standard boundary condition is imposed that the buried fields merge continuously with an external vacuum field. This boundary condition reads
\begin{equation}
\label{8}
 R\frac{\partial S}{\partial r}\,+\,S = 0
\end{equation}
for a dipole component. At the crust-core boundary  $r=\,r(b)$, it is assumed that the magnetic stream is conserved, e.g., $S=$ Constant. In such a case, the surface field $B_{\rm d}$ is increasing at the current epoch with a power$-$law form,
\begin{equation}
\label{9}
 B_{\rm d}(t)=  B_{\rm d}(0)\times \left(\frac{t}{1\, \rm yr}\right)^{\varepsilon},
\end{equation}
and is responsible for the spin-down evolution of the star including $n<3$, where $\varepsilon >0$ is the magnetic field index.

\subsection{Constraining the true age of the star}
As we know, when we investigate the long-term spin evolution of a pulsar, it is best to know the true age of the pulsar. Since $\tau_{\rm c}$ is a poor age approximation of a pulsar, its true age $t_{\rm age}$ can be estimated by the age of its associated supernova remanant\,(SNR)\,(Gao et al. 2017). It is supposed that PSR J1734$-$3333 is associated with a shell remnant G354.8$-$0.8\,(e.g., Ho 2012; Pavlovic et al. 2014). Since no X-ray emission was detected from the SNR's shell, the true age of PSR J1734$-$3333 cannot be estimated. In the previous study, a rather low age limit of $t_{\rm age}=t_{\rm SNR} >1.3$\,kyrs was adopted (e.g., Gourgouliatos \& Cumming 2015). Ho (2012) estimated $t_{\rm age }\sim$ 2.0 kyrs for the star by considering the size of G354.8$-$0.8 as 21 parsecs and remnant
expansion velocity $v_{\rm SNR}\sim 10^4$ km~s$^{-1}$; and considering the pulsar¡¯s distance away from the center of the remnant about 46 parsecs\,(Manchester et al. 2002) and pulsar space velocity $v_{\rm PSR}\sim$ 2000 km~s$^{-1}$, then obtained an age $t_{\rm age}\sim 23$ kyrs. However, if $n < 3$, then $\tau_{\rm c} < t_{\rm SNR}$, and a pulsar appear ¡°younger¡± than it is\,(Gao et al. 2016), the smaller the braking index is, the larger the disparity between $\tau_{\rm c}$ and $t_{\rm SNR}$. This required that $t_{\rm SNR}$ is larger than $\tau_{\rm c}$ for the pulsar\,($\tau_{\rm c}=8.13$\,kyrs). Recently, Pavlovic et al (2014) present new empirical radio surface-brightness-to diameter ($\Sigma-D$)relations for SNRs in our Galaxy, and estimated the  diameter $D\sim $34.8 parsecs and the distance $d\sim$ 6.3 kpc from flux-density 2.8 Jy for G354.8$-$0.8, which corresponds a small age of about 20 kyrs. In this work, the true age of the age is taken as 20-23\,kyrs.

\subsection{Constraining initial parameters}
If the magnetic field evolution of PSR J1640$-$4631 cannot be ignored, and the dipole braking still dominates, according to Blandford \& Romani (1988), the braking law of the pulsar is reformulated as
\begin{equation}
\label{10}
\dot{\nu}(t)=-\frac{8\pi^{2}R^{6}sin^{2}\alpha}{3Ic^{3}}B^{2}_{\rm d}(t)\nu^{3}.
\end{equation}
Integrating Equation (10) gives the spin frequency,
\begin{equation}
\label{11}
\nu^{-2}=\nu^{-2}_{0}+2\int^{t}_{0}\frac{8\pi^{2}R^{6}sin^{2}\alpha
}{3Ic^{3}}B^{2}_{\rm d}(t^{'})dt^{'},
\end{equation}
where $\nu_{0}$ is the initial spin frequency of the pulsar. Then, we get the spin period,
\begin{eqnarray}
\label{12}
&&P(t)=\left[P_{0}+\int^{t}_{0}\frac{12\pi^{2}R^{6}sin^{2}\alpha}{3Ic^{3}}
B^{2}_{\rm d}(t^{'})dt^{'}\right]^{\frac{1}{2}},\nonumber\\
&&=\left[P^{2}_{0}+\frac{16\pi^{2}R^{6}B^{2}_{\rm d}(0)sin^{2}\alpha}{3Ic^{3}}
\frac{(t/1{\rm yr})^{2\varepsilon+1}}{2\varepsilon+1}1\,{\rm yr}\right]^{\frac{1}{2}}.
\end{eqnarray}
Then we can represent the spin-down age of the star in the form of
\begin{equation}
\label{13}
\tau_{\rm c}=\frac{-\nu}{2\dot{\nu}}=\frac{P}{2\dot{P}}=\frac{K}{B^{2}_{\rm d}(t)}\int^{t}_{0}B^{2}_{\rm d}(t^{'})dt^{'},
\end{equation}
where $K=[1-(P_{0}/P)^{2}]^{-1}$. Since we have assumed $B_{\rm d}\propto t^{\varepsilon}$ (see in Equation (9)), then we obtain
\begin{equation}
\label{14}
\tau_{\rm c}\sim \frac{K}{2\varepsilon + 1}\cdot t.
\end{equation}
Combining Equation (14) with Equation (5), we obtain
\begin{equation}
\label{15}
n\sim 3-\frac{2\varepsilon K}{ 2\varepsilon + 1}.
\end{equation}
The initial spin period $P_{0}$ can be estimated by
\begin{equation}
\label{16}
P_{0}=P(1+\frac{1}{K})^{1/2},\,\,K=\frac{3-n}{2}+\frac{\tau_{\rm c}}{t},
\end{equation}
when we take $t=t_{\rm age}$. In the same way, the magnetic growth index $\varepsilon$ is determined by
\begin{equation}
\label{17}
\varepsilon =\frac{3-\,n}{2(n-3+ 2K)}.
\end{equation}
Inserting the values of $\tau_{\rm c}$, $n$ and $B_{\rm d}$ at the current age $t_{\rm age}=20-23$\,kyrs into Equations (16-17)
 fields $P_{0}=65-62$\,ms,\,$B_{\rm d}(0)=(1.6-0.4)\times 10^{8}$\,G and $\varepsilon \sim (1.4-1.28)$.
\subsection{The spin-down evolution }
Utilizing the differential method, we get the first-order
derivative of the spin period $\dot{P}(t)$,
\begin{eqnarray}
\label{18}
&&\dot{P}(t)=\frac{8\pi^{2}R^{6}sin^{2}\alpha B^{2}_{\rm d}(0)(t/1{\rm yr})^{2\varepsilon}}{3Ic^{3}}\times
 \nonumber\\
&&\left[P^{2}_{0}+\frac{16\pi^{2}R^{6}B^{2}_{\rm d}(0)sin^{2}\alpha}{3Ic^{3}}\cdot
\frac{(t/1{\rm yr})^{2\varepsilon+1}}{2\varepsilon+1}\cdot 1{\rm yr}\right]^{\frac{-1}{2}},
\end{eqnarray}
and the second-order derivative of the spin period $\ddot{P}(t)$,
\begin{eqnarray}
\label{19-equation}
&&\ddot{P}(t)=\frac{16\varepsilon\pi^{2}R^{6}sin^{2}\alpha B^{2}_{\rm d}(0)(t/1{\rm yr})^{2\varepsilon-1}}{3Ic^{3}1\,{\rm yr}}\times \nonumber\\
 &&\left[P^{2}_{0}+\frac{16\pi^{2}R^{6}B^{2}_{\rm d}(0)sin^{2}\alpha}{3Ic^{3}}
\frac{(t/1{\rm yr})^{2\varepsilon+1}}{2\varepsilon+1}1\, {\rm yr}\right]^{\frac{-1}{2}}\nonumber\\
&& -\left(\frac{8\pi^{2}R^{6}sin^{2}\alpha B^{2}_{\rm d}(0)(t/1{\rm yr})^{2\varepsilon}}{3Ic^{3}}\right)^{2}\times \nonumber\\
 &&\left[P^{2}_{0}+\frac{16\pi^{2}R^{6}B^{2}_{\rm d}(0)sin^{2}\alpha}{3Ic^{3}}\cdot
\frac{(t/1{\rm yr})^{2\varepsilon+1}}{2\varepsilon+1}\cdot 1{\rm yr}\right]^{\frac{-3}{2}}.
\end{eqnarray}
Inserting Equation (12), Equation (18) and Equation (19) into $n=2-\frac{P\ddot{P}}{\dot{P}^{2}}$, we have
\begin{eqnarray}
\label{20}
&&n=3-\frac{3Ic^{3}\varepsilon}{4\pi^{2}R^{6}sin^{2}\alpha B^{2}_{\rm d}(0)(t/1{\rm yr})^{2\varepsilon}t}\times \nonumber\\
&&\left[P^{2}_{0}+\frac{16\pi^{2}R^{6}B^{2}_{\rm d}(0)sin^{2}\alpha}{3Ic^{3}}
\frac{(t/1{\rm yr})^{2\varepsilon+1}}{2\varepsilon+1}1{\rm yr}\right].
\end{eqnarray}
In order to investigate the evolution of $n$, we plot the diagram of $n$ versus $t$ for the pulsar.
\begin{figure}[!htbp]
\centering
 \includegraphics[width=0.45\textwidth]{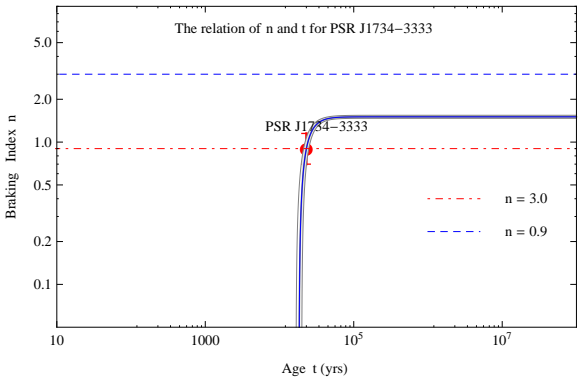}
\caption{Braking index as a function of $t$ for PSR J1344$-$3333.
The measured value of $n$ is shown with the red dot and the error bar denotes the possible range given by the uncertainty $\Delta n=0.2$ for the pulsar.}
\label{fig_1}
\end{figure}
In Figure 1, the blue full line  stands for the prediction of the dipole magnetic field growth model given in Equation (20), and the blue dashed line stands for $n=3$ predicted by the MDR model. The surrounding shaded region around the blue full line denotes the fit uncertainty.
The horizontal red dot-dashed line denotes $n=0.9$. Here and following, we adopt $\alpha=90^{\circ}$ and $I=10^{45}$\,g~cm$^{2}$ for PSR J1734$-$3333, corresponding to a NS mass of 1.4\,$M_{\rm sun}$ and the radius of $R=10^{6}$ cm\,(Gunn \& Ostriker 1969; Contopoulos et al. 2014). From Fig. 1, one can see that the growth of the dipole magnetic field causes a low braking index $n\,<\,3$. If $B_{\rm d}$ increases with a constant index of $\varepsilon=1.34$, the braking index $n$ will increase continually until it reaches a limit $n\sim 1.5$. Not that an evolving $\alpha$ can produce similar spin evolution behaviour to one with $\dot{B}$.  However evidence for a varying $\alpha$ is uncertain, as discussed in Guill\'{o}n et al. (2014) and Ho (2015).
\subsection{The relation of $\tau_{\rm c}$ and $t_{\rm age}$ }
Inserting Equation (12) and Equation (18) into $\tau_{\rm c}=P/2\dot{P}$, we get
\begin{eqnarray}
\label{21}
&&\tau_{\rm c}=\frac{3Ic^{3}}{16\pi^{2}R^{6} B^{2}_{\rm d}(0)(t/1{\rm yr})^{2\varepsilon}}\times
 \nonumber\\
&&\left[P^{2}_{0}+\frac{16\pi^{2}R^{6}B^{2}_{\rm d}(0)}{3Ic^{3}}\cdot
\frac{(t/1{\rm yr})^{2\varepsilon+1}}{2\varepsilon+1}\cdot 1{\rm yr}\right].
\end{eqnarray}
In Figure 2, we plot $\tau_{\rm c}$ versus $t$ for the pulsar.
\begin{figure}[!htbp]
\centering
 \includegraphics[width=0.45\textwidth]{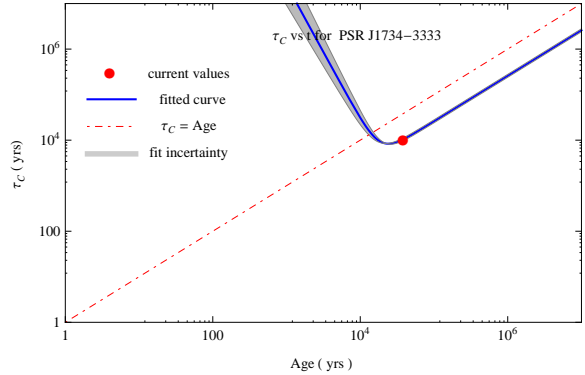}
\caption{The characteristic age as a function of $t$ for PSR J1734$-$3333. The measured value of $\tau_{\rm c}$ is shown with the red dot. }
\label{fig_2}
\end{figure}

In Figure 2, the blue full line stands for the prediction of diploe magnetic field growth model given by Equation (21), and the red dot-dashed line denotes $\tau_{\rm c}=t_{\rm age}$. One can see
that $\tau_{\rm c}$ decreases with the age at the earlier stage, then increases with the age at the latter stage. Suppose that, a growing $B_{\rm d}$ causes an increase in the braking torque $N$, which directly causes an increase in $n$ (from a negative to a positive value), but $n$ is always less than 3 predicted by MDR model.
\subsection{The evolution of the dipole magnetic field}
Using the constrained parameters of $t_{\rm age}$, $\tau_{\rm D}$ and $B_{\rm p}(0)$, we firstly estimate a mean magnetic field decay rate of the pulsar:\,$\Delta B_{\rm d}/\Delta t=(B_{\rm d}(t_{\rm age})-B_{\rm d}(0))/t_{\rm age}\approx (2.27-2.61)\times10^{9}$\,G\,yr$^{-1}$.
then make the diagram of $B_{\rm d}$ versus $t$ of PSR J1734$-$3333, shown as in Figure 3(a).

The decay rate $\dot{B}_{\rm d}$ of PSR J1734$-$3333 is also an important issue.  From Equation (9), we obtain the expression of $\dot{B}_{\rm d}$ and $t$ for the pulsar,
\begin{equation}
\label{21}
\frac{dB_{\rm d}(t)}{dt}=\frac{\varepsilon B_{\rm d}(0)(t/1\,{\rm yr})^{\varepsilon -1}}{1\,{\rm yr}}.
\end{equation}
The magnetic field decay rate $\dot{B}_{\rm d}$ of the pulsar increases with $t$, as shown in Figure 3(b). Here the term ``$\dot{B}_{\rm d}$ increase'' refers to an increase in the magnitude of $|dB_{\rm d}/dt|$. When $t=t_{\rm age}$, we get the present value of $dB_{\rm d}/dt\simeq (1.3\times10^{9}-1.2\times10^{10})$ G yr$^{-1}$. If we insert parameter groups $B_{\rm d}(t_{\rm age})\sim 5.22\times10^{13}$\,G, $dB_{\rm d}/dt\simeq 1.3\times10^{9}$ G yr$^{-1}$, $t_{\rm age}=20$\,kyrs and $\varepsilon=1.40$\, (or parameter groups $B_{\rm d}(t_{\rm age})\sim 5.22\times10^{13}$\,G, $dB_{\rm d}/dt\simeq 1.2\times10^{10}$ G yr$^{-1}$, $t_{\rm age}=23$\,kyrs and $\varepsilon=1.28$) into Equation (5), then we obtain $n=0.9$, which is consistent with the measurement of $n$ for the pulsar.
\begin{figure}[th]
\centering
 \vspace{0.6cm}
\subfigure[]{
    \label{evsb:a} 
    \includegraphics[width=7.2cm]{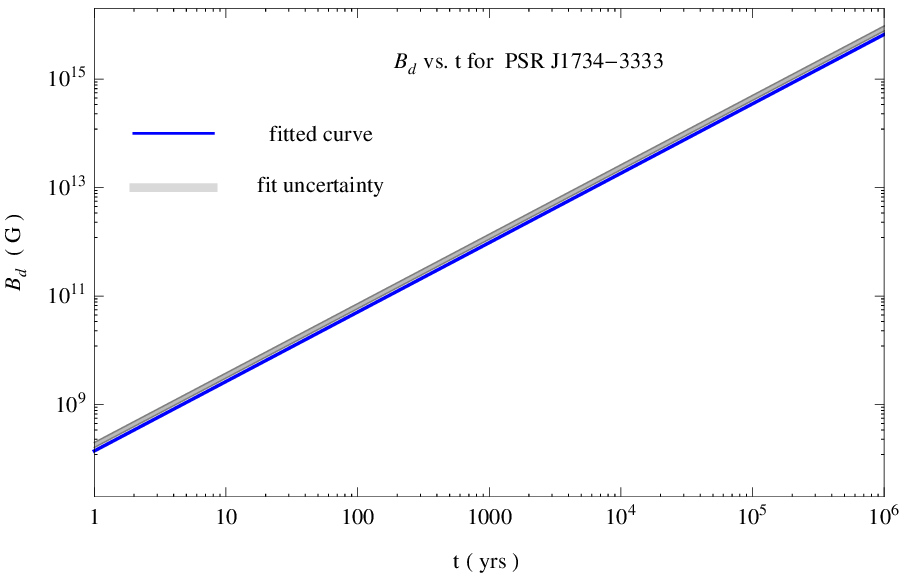}}
  \hspace{0.2mm}
  \subfigure[]{
    \label{evsb:b} 
    \includegraphics[width=7.2cm]{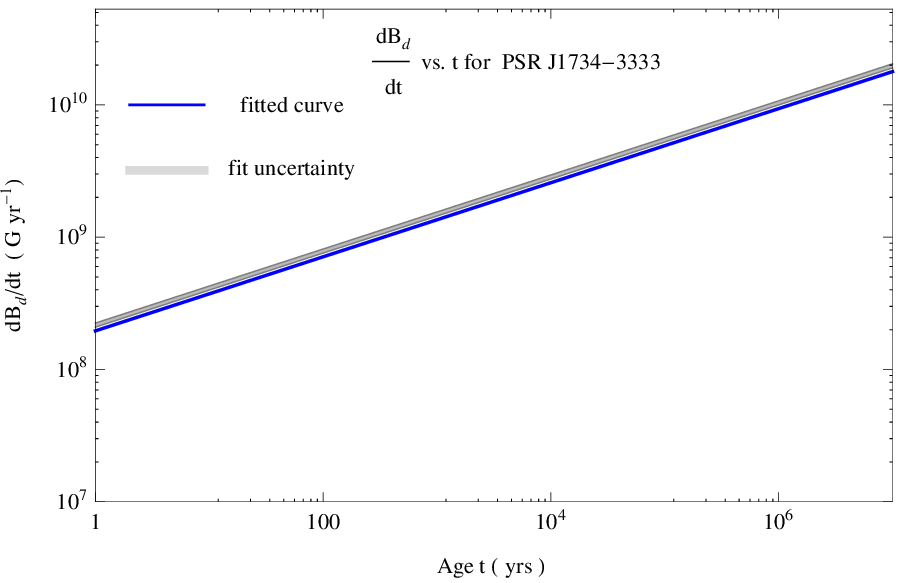}}
 \caption{Top, the relation between $B_{\rm d}$ and $t$ of PSR J1734$-$3333. Bottom, the relation between $\dot{B}_{\rm d}$ and $t$ of PSR J1734$-$3333.}
 \label{fig2}
 \end{figure}
 In the above figure, the blue full lines stand for the predictions of the dipole magnetic field growth model given by Equation (9) and Equation (21). The surrounding shaded regions around the blue full lines denote the fit uncertainty. If the surface dipole field of PSR J1734$-$3333  increase with the current growth index of $\varepsilon=$1.34(6), this pulsar will become a magnetar with $B_{\rm d}\sim 2.6(1)\times 10^{14}$\,G, $B_{\rm d}\sim 5.3(2)\times 10^{14}$\,G, $B_{\rm d}\sim 1.2(1)\times 10^{15}$\,G, after the next 50\,kyrs, 100\,kys, and 200\,kyrs, respectively. In addition, we show the long-term rotational evolution of the pulsar in Figure 4.
 \begin{figure}[!htbp]
\centering
 \includegraphics[width=0.45\textwidth]{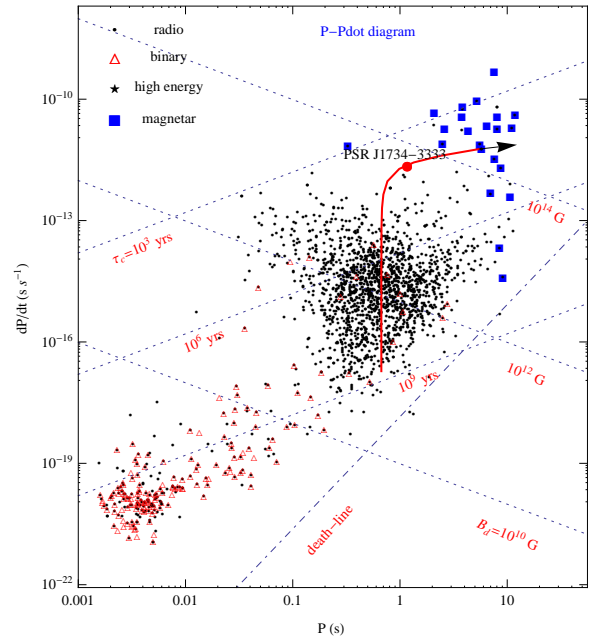}
\caption{Long term rotational evolution of PSR J1734$-$3333 dominated by dipole field growth. The red solid circle denotes the observations of the pulsar. }
\label{fig_4}
\end{figure}

 Here the field growth index $\varepsilon$ her is equivalent to an average field growth index. When the surface dipole field $B_{\rm d}$ approaches to the internal field $B_{\rm in}$ at the saturated region, $\varepsilon$ of PSR J1734$-$3333 may change\,(may be smaller). A similar situation will happen in the other NSs. Gourgouliatos \& Cumming (2014) investigated magnetic field evolution in the NS crust due to Hall drift as an explanation for observed braking indices $n<3$ of pulsars including PSR J1734$-$3333, they pointed that rapid interior cooling after the next 100 kyrs will stop the surface field growth. In that case, its surface dipole field  will decrease, because th surface neutrino and photon emission cause the NS¡¯s  cooling, which in turn speeds up the decay of $B_{\rm d}$, mainly through Ohmic diffusion. However we mainly focus on the scenario of the surface magnetic field growth for PSR J1734$-$3333 within one~several hundred kyrs, the scenario of the surface magnetic field decay will be beyond the scope of this work.

\section{ Discussion and conclusions}
In this work, we present a possible interpretation for very small braking index of PSR J1734$-$3333, which challenges the current theories of braking mechanisms in pulsars, and estimate some parameters including the initial spin-period and the initial dipole magnetic field strength of the star.  According to our suggestions, this pulsar could be  born with a superhigh internal magnetic field $\sim10^{14}-10^{16}$ G, and could undergo a supercritical accretion soon after its formation in a supernova. This strong magnetic field has been buried under the surface, and is relaxing out of the surface at present due to Ohmic diffusion. Keep the current field-growth index $\varepsilon=1.34(6)$, the surface dipole field would reach the maximum of the internal magnetic field strength in a few hundred thousand years, which implies that this pulsar is a potential magnetar.

The maximum uncertainty of field-growth index could be from the age estimation of G354.8-0.8. Due to lack of X-ray emission and the accurate measurements of distance and radius, we can not present an accurate estimate of true age for the SNR.  We expect that  future observations will provide us an appropriate age range. Thus, the initial parameters in this work will be modified substantially, according to the observations.
\acknowledgements
This work was supported by by National Basic Research Program of China grants 973 Programs 2015CB857100.
This work is also supported by the West Light
Foundation of CAS through grants XBBS-2014-23, XBBS-2014-22 and 2172201302, Chinese National
Science Foundation through grants No.11673056, 11273051,
11373006 and 11133001, 11173042, the Strategic Priority Research
Program of CAS through No. XDB09000000 and XDB23000000, and by a research fund from the Qinglan project of Jiangsu Province.

\end{document}